\documentclass[doublecol]{epl2}

\usepackage{amssymb,amsmath}
\usepackage{graphicx}

\begin{document}

\title{Disorder correlations at smeared phase transitions}

\author{Christopher Svoboda \and David Nozadze \and Fawaz Hrahsheh \and Thomas Vojta}
\shortauthor{C. Svoboda, D. Nozadze, F. Hrahsheh and T. Vojta}
\institute{Department of Physics, Missouri University of Science and Technology, Rolla, MO 65409, USA}

\abstract{
We investigate the influence of spatial disorder correlations on smeared phase transitions, taking the magnetic quantum
phase transition in an itinerant magnet as an example. We find that even short-range correlations can have a dramatic effect
and qualitatively change the behavior of observable quantities compared to the uncorrelated case.
This is in marked contrast to conventional critical points, at which short-range correlated disorder and uncorrelated
disorder lead to the same critical behavior.
We develop an optimal fluctuation theory of the quantum phase transition in the presence of correlated
disorder, and we illustrate the results by computer simulations. As an experimental application, we discuss
the ferromagnetic quantum phase transition in Sr$_{1-x}$Ca$_x$RuO$_3$.
}

\date{\today}

\maketitle

\section{Introduction}

Quenched disorder has various important consequences in condensed matter.
For example, disorder can change the universality class of a critical
point \cite{HarrisLubensky74,GrinsteinLuther76}
or even change the order of a phase transition \cite{ImryWortis79,AizenmanWehr89,HuiBerker89}.

In theoretical studies, the disorder is often assumed
to be uncorrelated in space even though many sample preparation techniques will produce
some degree of correlations between the impurities and defects. As long as the correlations are
short-ranged, i.e., characterized by a finite correlation length $\xi_{\rm dis}$, this assumption is usually
justified  if one is interested in the universal properties of critical points.
(There are exceptions for special, fine-tuned local correlations \cite{HLVV11}).
The reason why short-range correlated disorder leads to the same behavior as uncorrelated disorder
can be easily understood within the renormalization
group framework. Under repeated coarse graining, a nonzero disorder correlation length $\xi_{\rm dis}$ decreases
without limit.
The disorder thus becomes effectively uncorrelated on the large length scales that determine the
critical behavior.

A formal version of this argument follows from the Harris criterion \cite{Harris74}.
It states that a clean critical point is stable against weak \emph{uncorrelated} disorder
if its correlation length critical exponent $\nu$ fulfills the inequality $d\nu > 2$
where $d$ is the space dimensionality. If the inequality is violated, the disorder is relevant and
changes the critical behavior. According to Weinrib and Halperin \cite{WeinribHalperin83},
\emph{spatially correlated} disorder leads to the same inequality as long as its correlations
decay faster than $r^{-d}$ with distance $r$. Thus, short-range correlated disorder and
uncorrelated disorder have the same effect on the stability of a clean critical point.

\begin{figure}
\includegraphics[width=8cm]{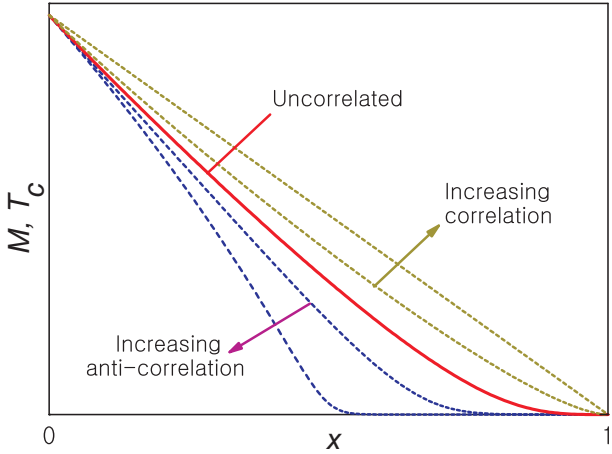}
\caption{(Color online) Schematic of the zero-temperature magnetization-composition curve ($M$ vs $x$) and
the finite-temperature phase boundary ($T_c$ vs $x$) at a smeared
quantum phase transition in a random binary alloy A$_{1-x}$B$_x$. The cases of uncorrelated, correlated, and anti-correlated disorder
are contrasted.}
\label{fig:overview}
\end{figure}

In this Letter, we demonstrate that spatial disorder correlations are much more important at
\emph{smeared} phase transitions, a broad class of classical and quantum phase transitions
characterized by a gradual, spatially inhomogeneous onset of the ordered phase \cite{2006_Vojta_JPhysA}.
Specifically, we show that short-range correlated disorder and uncorrelated disorder
lead to qualitatively different behaviors.  The disorder correlations do not only influence quantities usually considered
non-universal such as the location of the phase boundary, they also change the functional
dependence of the order parameter and other quantities on the tuning parameters of the transition,
as indicated in Fig.\ \ref{fig:overview}.
We propose that this mechanism may be responsible for the unusually wide variations reported in the literature
on the properties of the ferromagnetic quantum phase transition (QPT) in Sr$_{1-x}$Ca$_x$RuO$_3$.

In the following, we sketch the derivation of our theory, compute observables,
and illustrate them by simulations. We also discuss the generality of our findings,
and we compare them to experiment.

\section{Smeared quantum phase transition}

For definiteness, we consider a magnetic QPT in a metallic system with Ising order
parameter symmetry. In the absence of quenched disorder, the Landau-Ginzburg-Wilson
free energy functional of this transition is given by \cite{Hertz76,Millis93}
\begin{equation}
S=\int dy dz ~\psi(y)\Gamma(y,z)\psi(z)+ u \int dy~\psi^{4}(y)\,,
\label{eq:clean-action}
\end{equation}
where $\psi$ is the order parameter field, $y\equiv(\mathbf{y},\tau)$ comprises
$d$-dimensional spatial position $\mathbf{y}$ and imaginary time $\tau$, the integration means
$\int dy\equiv\int d\mathbf{y}\int_{0}^{1/T}{\rm d}\tau$, and $u$
is the standard quartic coefficient. The Fourier transform of the Gaussian vertex $\Gamma(y,z)$ reads
\begin{equation}
\Gamma(\mathbf{q},\omega_{n})=r+\xi_{0}^{2}\mathbf{q}^{2}+\gamma_0(\mathbf{q})\left|\omega_{n}\right|~.
\label{eq:bare_Gamma}
\end{equation}
Here, $r$ is the distance from criticality,\footnote{Strictly, one needs to distinguish the bare distance
from criticality that appears in (\ref{eq:bare_Gamma}) from the renormalized one that measures the distance
from the true critical point. We suppress this difference because it is unimportant for our purposes.}
 $\xi_{0}$ is a
microscopic length, and $\omega_{n}$ is a Matsubara frequency. The dynamical part of
$\Gamma(\mathbf{q},\omega_{n})$ is proportional to $|\omega_{n}|$. This reflects the Landau damping of
the order parameter fluctuations by gapless electronic excitations in a metallic system.
The coefficient $\gamma_0(\mathbf{q})$ is $\mathbf{q}$-independent for an antiferromagnetic
transition but proportional to $1/|\mathbf{q}|$ or $1/|\mathbf{q}|^2$ for ballistic and diffusive ferromagnets,
respectively.

We now consider a random binary alloy A$_{1-x}$B$_x$
consisting of two materials A and B. Pure substance B has a non-magnetic ground-state, implying
a positive distance from quantum criticality, $r_{\rm{B}}>0$. Substance A has a
magnetically ordered ground state with $r_{\rm{A}}<0$. By randomly substituting B atoms for A atoms,
one can drive the system through a QPT from a magnetic to a nonmagnetic
ground state.

Due to statistical fluctuations, the distribution of A and B atoms in the alloy will not be
spatially uniform. Some regions may contain significantly more A atoms than the average. If the local
A-concentration is sufficiently high, such regions will be locally in the magnetic phase even
if the bulk system is nonmagnetic. Because the magnetic fluctuations are overdamped,
the {quantum} dynamics of sufficiently large such locally magnetic spatial regions completely
freezes (for Ising symmetry \cite{2001_Millis_PRL}). {At zero temperature}, these rare regions thus
develop static magnetic order independently of each other. This destroys the sharp QPT by smearing
\cite{2003_Vojta_PRL,2006_Vojta_JPhysA,2010_Vojta_JLTPhys} and is manifest in a
pronounced tail in the {zero-temperature} magnetization-composition curve \cite{2011_Hrahsheh_PRB}.

{
At any nonzero temperature, the static magnetic order on individual, independent rare regions is destroyed because
they can fluctuate via thermal excitations. Therefore, a finite interaction between the rare regions of the order of
the thermal energy is necessary to align them. This restores a conventional sharp phase transition at any nonzero
temperature.
However, the smeared character of the underlying QPT leads an unusual concentration dependence of the critical temperature $T_c$
which displays a tail towards large $x$ \cite{2003_Vojta_PRL,2011_Hrahsheh_PRB}.
}

The effects of disorder correlations can be easily understood at a qualitative level. For positive correlations,
like atoms tend to cluster. This increases, at fixed composition, the probability of finding large A-rich regions
compared to the uncorrelated  case. The tail of magnetization-composition curve therefore becomes larger
(see Fig.\ \ref{fig:overview}). In contrast, like atoms repel each other in the case of negative correlations
(anti-correlations).
This decreases the probability of finding large A-rich regions and thus suppresses the tail.

\section{Optimal fluctuation theory}

To quantify the influence of the disorder correlations, we now develop an
optimal fluctuation theory \cite{2003_Vojta_PRL,2011_Hrahsheh_PRB}.
We focus on the ``tail'' of the smeared transition (large $x$) where a few rare regions show
magnetic order but their interactions are weak because they are far apart.

We roughly estimate the transition point in the alloy  $\rm{A}_{1-\emph{x}}\rm{B}_{\emph{x}}$,
by setting the average distance from criticality to zero,
$
r_{\rm{av}}=(1-x)r_{\rm{A}}+x r_{\rm{B}}=0.
$
This defines the critical composition in ``average-potential'' approximation,
\begin{equation}
x^0_c=-r_{\rm{A}} /(r_{\rm{B}}-r_{\rm{A}})\label{XO}\,.
\end{equation}
For compositions $x>x_c^0$, static magnetic order can only develop on rare, atypical
spatial regions with a higher than average A-concentration. Specifically,
a single A-rich rare region of linear size $L_{\rm{RR}}$
can show magnetic order, if the local concentration $x_{\rm{loc}}$ of
B atoms is below some critical value  $x_c$. Because the rare region
has a finite size, the critical concentration is shifted  from the bulk value $x^0_c.$
According to finite-size scaling \cite{1983_Barber_Academic,1988_Cardy_NH}
\begin{eqnarray}
x_c(L_{\rm{RR}})=x^0_c-DL^{-\phi}_{\rm{RR}}\label{XC}\,,
\end{eqnarray}
where $\phi$ is the finite-size shift exponent and $D$ is a non-universal constant.
In a three-dimensional itinerant magnet, $\phi$ takes the mean-field value
of 2 because the clean transition is above its upper critical dimension.
As $x_c(L_{\rm{RR}})$ must be positive, a rare region must be larger than
 $L_{\rm{min}}=(D/x^0_c)^{1/\phi}$ to show magnetic order.

In the tail of the smeared transition, the magnetically ordered  rare regions are far apart
and interact only weakly. To find the total magnetization $M$  one can thus
simply sum over all magnetically ordered rare regions. This gives
\begin{equation}
M\sim\int_{L_{\rm{min}}}^{\infty}\! dL_{RR} \int_{0}^{x_{c}(L_{\rm{RR}})} dx_{\rm{loc}} P(N,x_{\rm{loc}})
m(N,x_{\rm{loc}})\label{MInt}\,,
\end{equation}
where $P(N,x_{\rm{loc}})$ is the probability for finding a region of
$N$ sites and local composition $x_{\rm loc}$ (i.e., a region containing $N_B=Nx_{\rm{loc}}$ atoms of type B),
and  $m(N,x_{\rm{loc}})$ is its magnetization

Let us analyze the spatial distribution of atoms in the sample to determine the probability $P(N,x_{\rm{loc}})$.
Specifically, let us assume that the random positions of the A and B atoms are positively
correlated such that like atoms form clusters of typical correlation volume (number of lattice sites)
$V_{\rm dis} \approx 1 + a \xi_{\rm dis}^d$ where $\xi_{\rm dis}$ is the disorder correlation length and
 $a$ is a geometric prefactor. The probabilities for finding A and B clusters in the sample are $1-x$ and $x$, respectively.
The number $n_{\rm{cl}}$ of correlation clusters contained in a large spatial region of $N$ sites ($N\gg V_{\rm dis}$)
is approximately
\begin{equation}
n_{\rm{cl}} \approx N/V_{\rm dis} =N/(1+a \xi^d_{\rm{dis}})
 ~.
\end{equation}

The probability $P(N,x_{\rm{loc}})$ for finding a region of $N$ sites and local composition $x_{\rm loc}$
is therefore equal to the probability $P_{\rm{clus}}(n_{\rm{cl}},n_{\rm{B}})$
for finding $n_{\rm{B}} = x n_{\rm{cl}}$ clusters of B atoms among all the $n_{\rm{cl}}$ clusters
contained in the region. It can be modeled by a binomial distribution
\begin{equation}
P_{\rm{clus}}(n_{\rm{cl}},n_{\rm{B}})=\binom {n_{\rm{cl}}} {n_{\rm{B}}}(1-x)^{n_{\rm{cl}}-n_{\rm{B}}} x^{n_{\rm{B}}}\label{PClusBin}\,.
\end{equation}
We now distinguish two cases, (i) the regime where $x$ is not much larger than $x^0_{c}$,
and (ii) the far tail of transition at $x\to 1.$

(i) If $x$ is just slightly larger than $x_c^0$, rare regions
are large and the probability (\ref{PClusBin}) can be approximated by a Gaussian
\begin{equation}
P_{\rm{clus}}\approx \frac{1}{\sqrt{2 \pi  x(1-x)/n_{\rm{cl}}}}
         \exp\left[{-n_{\rm{cl}}\frac{(x_{\rm{loc}}-x)^2}{2x(1-x)}}\label{PclusGaus}\right] \,.
\end{equation}
We estimate the integral (\ref{MInt}) in saddle point approximation. Neglecting subleading
contributions from $m(N,x_{\rm{loc}}),$ we find that rare regions of size
$
L^{\phi}_{\rm{RR}}=D(2\phi-d)/[d(x-x^0_c)]
$
and composition $x_c(L_{\rm RR})$ dominate the integral. The resulting $M(x)$ dependence  reads
\begin{equation}
M\sim \exp \left[-\frac{C}{(1+a\xi^d_{\rm{dis}})}\frac{(x-x^0_c)^{2-d/\phi}}{x(1-x)}\right]\label{MExp}\,,
\end{equation}
where $C=2(D/d)^{d/\phi}(2\phi-d)^{d/\phi-2}\phi^2$ is a non-universal constant. In this regime, varying the
disorder correlation length thus modifies the non-universal prefactor of the exponential dependence of $M$ on $x$.

(ii) An even more striking effect occurs in the tail of the transition for $x\to 1.$
As rare regions cannot be large in this regime, the binomial distribution (\ref{PClusBin})
cannot be approximated by a Gaussian. However, within saddle point approximation,
the integral (\ref{MInt}) is dominated by rare regions  containing
only A atoms and having the minimum size permitting local order.
Inserting $L_{RR}=L_{\rm{min}}=(D/x^0_c)^{1/\phi}$ and $x_{\rm{loc}}=0$
into (\ref{MInt}), we find that the
composition dependence of the magnetization is given by the
power law,
\begin{equation}
M\sim (1-x)^\beta \qquad (x\to 1)\label{MPow}\,,
\end{equation}
with $\beta={a L^d_{\rm{min}}/{(1+a\xi^d_{\rm{dis}})}}$.
In this regime, the disorder correlations thus modify the seeming critical exponent of
the order parameter.  The exponent value is  given by the minimum number of correlation clusters necessary to form
a magnetically ordered rare region. The results for uncorrelated disorder \cite{2011_Hrahsheh_PRB} are recovered
by substituting $\xi_{\rm{dis}}=0$ into  (\ref{MExp}) and (\ref{MPow}).

So far we have assumed that a typical disorder correlation cluster of A atoms is smaller than the minimum
rare region size required for magnetic order.
For larger disorder correlation length $\xi_{\rm{dis}}\geq L_{\rm{min}},$ a single correlation cluster
is already large enough to order magnetically. As a result, (almost) all A atoms contribute to the
total magnetization. Correspondingly, the composition dependence of the order parameter is given by
\begin{equation}
M\sim (1-x)\label{MPow2}\,.
\end{equation}
To combine the power laws  (\ref{MPow}) and (\ref{MPow2}) for different ranges of $\xi_{\rm{dis}}$,
we construct the heuristic formula
\begin{equation}
\beta=(aL^d_{\rm{min}}+a\xi^d_{\rm{dis}})/(1+a\xi^d_{\rm{dis}}) \label{Exponent}\,
\end{equation}
which can be used to fit experimental data or simulation results.

{
Other observables such as the finite-temperature phase boundary can be found in similar fashion.
As discussed above, at $T\ne 0$, individual rare regions do not develop a static
magnetization. Instead, global magnetic order arises via a conventional (sharp) phase transition
at some transition temperature $T_{c}$ which can be estimated from the condition that the
interaction energy between the rare regions is of the order of the thermal energy.
To determine the interaction energy, we note that in a metallic magnet, the rare-regions
are coupled by an RKKY interaction which falls off as $r^{-d}$ with distance
$r$.
}
As the typical distance between neighboring rare regions behaves as
$r \sim M^{-1/d}$ \cite{2003_Vojta_PRL}, the composition dependence of the
critical temperature is analogous to that of the magnetization.
In particular,
\begin{equation}
T_c(x)\sim (1-x)^{\beta}\label{Tc}
\end{equation}
in the tail of the smeared transition, $x\to 1$.

\section{Simulations}

We now verify and illustrate the theoretical predictions  by performing computer simulations of a toy model
\cite{2003_Vojta_PRL,Vojta03b}.
Its Hamiltonian is motivated by the so-called quantum-to-classical mapping \cite{Sachdev_book99}
which relates a quantum phase transitions in $d$ space dimensions to a classical transition in $d+1$ dimensions.
The extra space dimension corresponds to imaginary time in the quantum problem.
Consequently, we consider a (3+1)-dimensional classical Ising model on a hypercubic lattice with three space
dimensions and a single imaginary time-like dimension. The interaction in the time-like direction
is long-ranged as the $|\omega_n|$ frequency dependence in (\ref{eq:bare_Gamma}) corresponds to a
$1/\tau^2$ in imaginary time.
{
In the toy model, we replace this interaction by an infinite-range interaction in time direction,
both on the same site and between spatial neighbors.\footnote{{Even though the bare action
(\ref{eq:clean-action}, \ref{eq:bare_Gamma}) does not have an
interaction between spatial neighbors at different imaginary times $\tau$, such a coupling will
be generated in perturbation theory (or under RG) from the short-range spatial interaction and
the long-range interaction in time.}}
}
This correctly reproduces the smeared character of the phase transition due to static magnetic order on the
rare regions.
The Hamiltonian of the toy model takes the
form
\begin{equation}\label{sim_hamil}
H = -\frac{1}{L_{\tau}} \sum_{\left< \mathbf{y}, \mathbf{z} \right>, \tau, \tau'} J_0 S_{\mathbf{y}, \tau} S_{\mathbf{z}, \tau'} - \frac{1}{L_{\tau}} \sum_{ \mathbf{y}, \tau, \tau'} J_{\mathbf{y}} S_{\mathbf{y}, \tau} S_{\mathbf{y}, \tau'}~,
\end{equation}
where $\mathbf{y}$ and $\mathbf{z}$ are space coordinates, $\tau$ is the time-like coordinate, and $S_{\mathbf{y},\tau}=\pm 1$.  $L_{\tau}$ is the system size in time and $\left< \mathbf{y}, \mathbf{z} \right>$ denotes pairs of nearest neighbors in space.  $J_{\mathbf{y}}$ is a binary random variable
whose value, $J_h$ or $J_l$, is determined by the
type of atom on lattice site $\mathbf{y}$. The values at different sites $\mathbf{y}$ and $\mathbf{z}$
are \emph{not} independent,
they are correlated according to some correlation function
$\mathcal{C} (\mathbf{y} - \mathbf{z})$.
The average concentrations of $J_h$-sites and $J_l$-sites are $1-x$ and $x$, respectively.

Treating the time-like dimension within mean-field theory, which is exact because of the infinite range of the interactions,
a set of coupled nonlinear equations emerge for the local magnetizations $m_{\mathbf{y}} = (1/L_{\tau}) \sum_{\tau} S_{\mathbf{y}, \tau}$,
\begin{equation}\label{coup_mean_field}
m_{\mathbf{y}} = \tanh{ \frac 1 {T_{cl}} \left( J_{\mathbf{y}} m_{\mathbf{y}} + \textstyle{\sum}_{\mathbf{z}} J_0 m_{\mathbf{z}} + h \right)}~.
\end{equation}
Here, the $\mathbf{z}$-sum is over the nearest neighbors of site $\mathbf{y}$, and $h$ is a tiny symmetry-breaking magnetic field.
According to the quantum-to-classical mapping, the classical temperature $T_{cl}$ is not related to the physical temperature
of the underlying quantum system (which is encoded in $L_\tau$) but rather some quantum control parameter that tunes the distance from the quantum phase transition.

The local mean-field equations (\ref{coup_mean_field}) can be solved efficiently in a self-consistency cycle.
In the two clean limits with either $J_{\mathbf{y}} = J_h$ or $J_{\mathbf{y}} = J_l$ for all $\mathbf{y}$,
the phase transition occurs at $T_h = J_h + 6J_0$ and $T_l = J_l + 6J_0$, respectively.  We choose a classical temperature between $T_h$ and $T_l$ and
control the transition by changing the composition $x$.


To generate the correlated binary random variables representing the site occupations, a version of the Fou\-rier-filtering method \cite{1996_Makse_PRE}
is implemented.  This method starts from uncorrelated Gaussian random numbers $u_{\mathbf{y}}$ and turns them into correlated Gaussian
random numbers $v_{\mathbf{y}}$ characterized by some correlation function
$\mathcal{C} (\mathbf{r})$.
This is achieved by transforming the Fourier components $\tilde{u}_{\mathbf{q}}$ of the uncorrelated random numbers according to
\begin{equation}\label{ffm_def}
{\tilde{v}}_{\textbf{q}} = \big[ \tilde{\mathcal{C}}(\textbf{q}) \big]^{\frac{1}{2}} \tilde{u}_{\textbf{q}},
\end{equation}
where $\tilde{\mathcal{C}}(\textbf{q})$ is the Fourier transform of $\mathcal{C}(\mathbf{r})$.
The $v_{\mathbf{y}}$ then undergo binary projection to determine the occupation of site $\mathbf{y}$; the site
 is occupied by atom A if $v_{\mathbf{y}}$ is greater than a composition-dependent threshold and by atom B if
$v_{\mathbf{y}}$ is less than the threshold.

\begin{figure}
\includegraphics[width=\columnwidth]{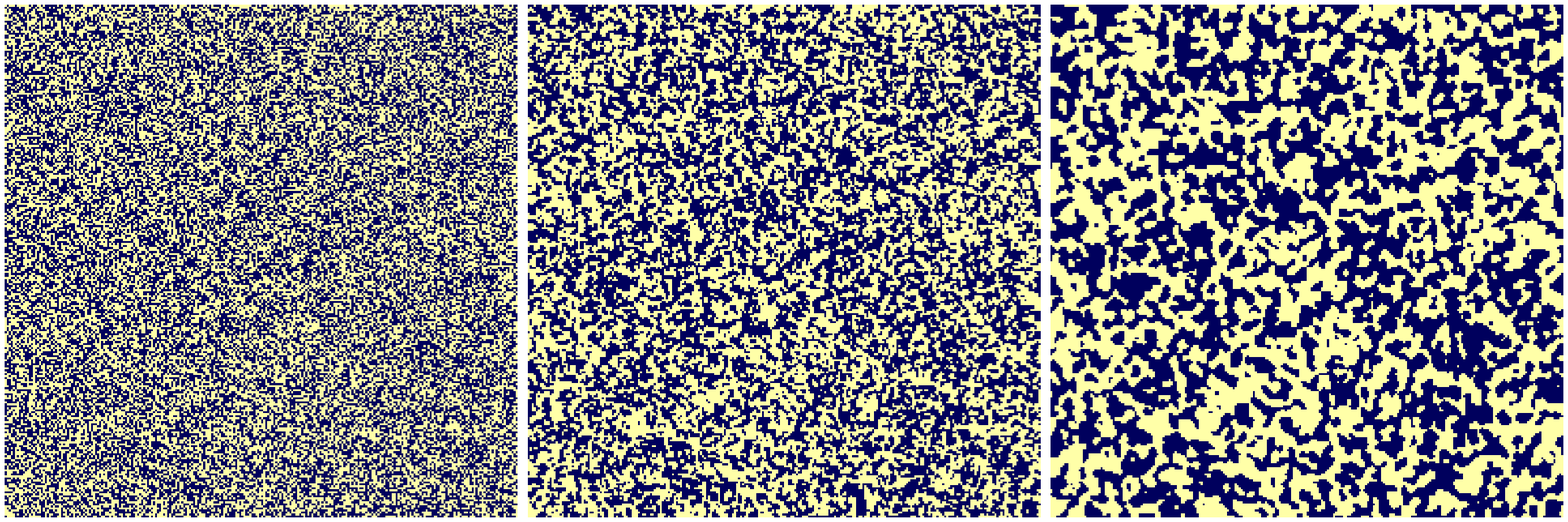}
\caption{(Color online) Examples of the atom distribution in a plane of 256$^2$ sites  for several values of the disorder
correlation length $\xi_{\rm dis} = 0, 1.0, 2.0$ from left to right ($x=0.5$).}
\label{vojta_fig1}
\end{figure}

{In the majority of our calculations, we focus on attractive} short-range disorder correlations of the form
$\mathcal{C}(\mathbf{r}) = \exp\left({-r^2/{2\xi_{\rm{dis}}^2}} \right)$.
Figure \ref{vojta_fig1} shows examples of the resulting atom distributions for
several values of the disorder correlation length $\xi_{\rm dis}$.
The formation of clusters of like atoms is clearly visible.

\begin{figure}
 \includegraphics[width=\columnwidth]{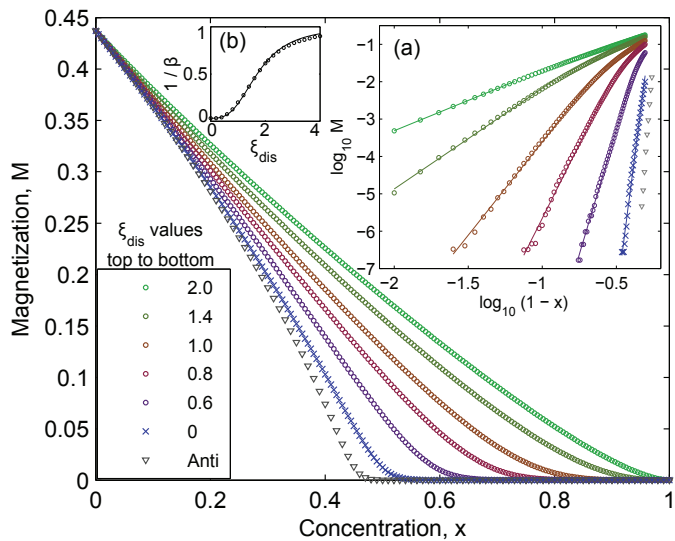}
 \caption{(Color online) Magnetization $M$ vs.\ composition $x$ for several values of the disorder correlation length $\xi_{\rm{dis}}$ using
 one disorder realization of $256^3$ sites, $J_h = 20$, $J_l = 8$, $J_0 = 1$, $T_{cl} = 24.25$, and $h = 10^{-10}$.
 {Also shown is one curve for the case of anti-correlations (128$^3$ sites), for details see text.} Inset (a): log-log plot of $M$ vs. $(1-x)$
 confirming the power-law behavior in the tail of the smeared transition. The tail exponent $\beta$ shown in inset (b) agrees very well
 with (\ref{Exponent}) as shown by the solid fit line.}
 \label{fig:m-T}
\end{figure}

We now discuss the results of the mean-field equations
(\ref{coup_mean_field}). Figure \ref{fig:m-T} presents the total magnetization $M$ as function of composition $x$  for several values
of $\xi_{\rm{dis}}$ with all other parameters held constant.
At a given composition $x$, the magnetization $M$ increases significantly even for small $\xi_{\rm{dis}}$ of the order of the lattice constant.
Moreover, the seeming transition point (at which $M$ appears to reach 0) rapidly moves towards larger compositions, almost reaching
$x=1$ for a correlation length $\xi_{\rm{dis}}=2$. Inset (a) of Fig.\ \ref{fig:m-T} shows a plot of
$\log M$ versus $\log (1-x)$ confirming the power-law behavior (\ref{MPow}) in the tail of the transition.  The dependence on $\xi_{\rm{dis}}$ of
the exponents $\beta$ extracted from these power laws is analyzed in inset (b) of Fig.\ \ref{fig:m-T}. It can be fitted well
with the heuristic formula (\ref{Exponent}).

{
In addition to the attractive (positive) correlations, we now briefly consider the case of anti-correlations
(like atoms repel each other). We model the anti-correlations by a correlation function having values
$\mathcal{C}(0)=1$, $\mathcal{C}(\mathbf{r})=-c$ for nearest neighbors, and $\mathcal{C}(\mathbf{r})=0$ otherwise.
The positive constant $c$ controls the strength of the anti-correlations.
A characteristic magnetization-composition curve for such anti-correlated disorder (with $c=1/6$) is included
in Fig.\ \ref{fig:m-T}. The data show that the magnetization is reduced compared to the uncorrelated case, and the
tail becomes less pronounced. Analogous simulations using different values of $c$ show that this effect increases
with increasing strength of the anti-correlations, as indicated in Fig.\ \ref{fig:overview}.

}

\section{Conclusions}

In summary, we have studied the effects of spatially correlated disorder on smeared phase transitions. We have found that
even short-range disorder correlations (extending over just a few lattice constants) lead to qualitative modifications of
the behavior at smeared transitions compared to the uncorrelated case, including changes in the exponents that characterize the order parameter and
the critical temperature. In other words, systems with uncorrelated disorder and with short-range correlated disorder
behave differently

This is in marked contrast to critical points, at which uncorrelated disorder and short-range correlated disorder lead to the same
critical behavior. (Long-range correlations do change the critical behavior \cite{WeinribHalperin83,RiegerIgloi99}.)
What causes this difference between critical points and smeared transitions? The reason is that
critical behavior emerges in the limit of infinitely large length scales while smeared transitions are governed by a
finite length scale, viz., the minimum size of ordered rare regions. This renders the renormalization group arguments
underlying the generalized Harris criterion \cite{Harris74,WeinribHalperin83} inapplicable.

{The majority of our calculations are for the case of like atoms attracting each other.
For these positive correlations, large locally ordered rare regions can form more easily than in
the uncorrelated case. Thus, the tail of the smeared transition is enhanced; and the phase boundary
as well as the magnetization curve move toward larger $x$ as indicated in Fig.\ \ref{fig:overview}.
We have also briefly considered the case of like atoms repulsing each other. These anti-correlations suppress the formation
of large locally ordered rare regions compared to the uncorrelated case. As a result, the phase boundary
and the magnetization curve will move toward smaller $x$.
}
In addition to short-range correlations, we have also studied long-range power-law correlations
which are interesting because they lead to a broad spectrum of cluster sizes.  Detailed results
will be published elsewhere \cite{Svoboda_etal_unpublished}.

\begin{figure}
\includegraphics[width=\columnwidth]{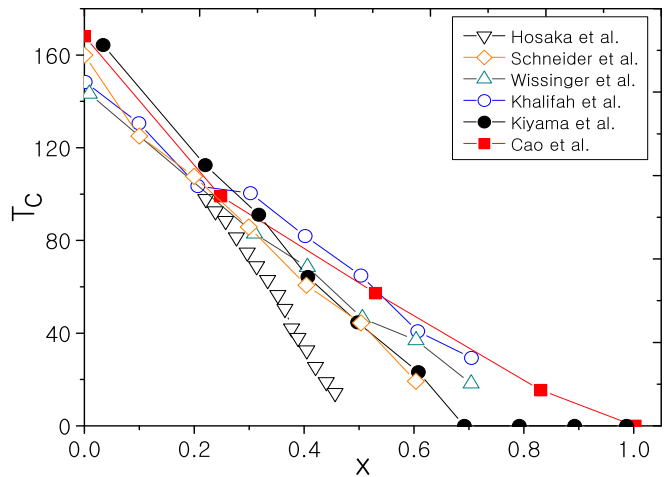}
  \caption{(Color online) Experimental temperature-composition phase diagrams of
 $\text{Sr}_{1-x}\text{Ca}_{x}\text{Ru}\text{O}_3$. Data from
 Hosaka et al.\ \cite{2008_Hosaka_ApplPhysEx}, Schneider et al.\ \cite{2010_Schneider_PSSB},
 Wissinger et al.\ \cite{2011_Wissinger_PRB}, and Khalifah et al.\ \cite{ 2004_Khalifah_PRB}
 are for thin films while those of Kiyama et al.\ \cite{1999_Kiyama_JPhysSJ}, and Cao et al.\
 \cite{1997_Cao_PRB} are for bulk samples. Published magnetization curves show similar variations.}
 \label{fig:exp}
\end{figure}

Turning to experiment, our results imply that smeared phase transitions are very sensitive
to slight short-range correlations in the spatial positions of impurities or defects.
In particular, an analysis of the data in terms of critical exponents will give values
that depend on these correlations.
We believe that a possible realization of the effects discussed in this paper can be found in
Sr$_{1-x}$Ca$_x$RuO$_3$. This well-studied material undergoes a ferromagnetic QPT as a
function of Ca concentration. Because Sr$_{1-x}$Ca$_x$RuO$_3$ is a metallic system with Ising spin symmetry,
the transition is expected to be smeared \cite{2003_Vojta_PRL}. Interestingly, the reported experimental
phase diagrams (see Fig.\ \ref{fig:exp}) and magnetization curves show unusually large
variations.
Not only does the apparent critical composition change between $x\approx 0.5$ and 1; the functional form
of the magnetization curves also varies.
Although part of these discrepancies may be due to the difference between film and bulk samples
\cite{2011_Wissinger_PRB}, large variations within each sample type remain. We propose that disorder
correlations, i.e., clustering or anti-clustering of like atoms may be responsible for at least part
of these variations.

Finally, we emphasize that even though we have considered the QPT in itinerant magnets
as an example, our theory is very general and should be applicable to all phase transitions smeared by
disorder including QPTs \cite{SchehrRieger06,SchehrRieger08,HoyosVojta08}, classical transitions in layered
systems \cite{SknepnekVojta04,Vojta03b} and non-equilibrium transitions \cite{Vojta04}


We thank I. Kezsmarki for helpful discussions. This work
has been supported in part by the NSF under grant No. DMR-
0906566.

\bibliographystyle{eplbib}
\bibliography{a}

\end{document}